\documentclass[11pt, a4paper]{article}
\pdfoutput=1
\usepackage{jheppub}
\usepackage{graphicx}
\usepackage{epsfig,amsmath,amsthm}
\usepackage{epstopdf}
\newcommand{\be}{\begin{equation}}
\newcommand{\ee}{\end{equation}}
\newcommand{\bea}{\begin{eqnarray}}
\newcommand{\eea}{\end{eqnarray}}
\def\bse{\begin{subequations}}
\def\ese{\end{subequations}}
\newcommand{\IR}{\mathbb{R}} 
 
\def\IZ{\relax\ifmmode\hbox{Z\kern-.4em Z}\else{Z\kern-.4em Z}\fi}

\newcommand{\non}{\nonumber \\}

\def\half{\frac{1}{2}} 

\def\del{{\partial}}

\def\al{\alpha}

\def\lam{\lambda}

\def\tal{\tilde{\alpha}}

\def\presub{\vspace{.5cm} \noindent}

\def\bi{\begin{itemize}} \def\ei{\end{itemize}}

\def\({\left(} \def\){\right)}
\def\[{\left[} \def\]{\right]}
\def\<{\left<} \def\>{\right>}

\title{Two-loop vacuum diagram through the Symmetries of Feynman Integrals method}

\author{Barak Kol  \\
{\it Racah Institute of Physics, Hebrew University, Jerusalem 91904, Israel} \\
{\tt barak.kol@mail.huji.ac.il}
}

\abstract{
The Symmetries of Feynman Integrals method (SFI) associates a natural Lie group with any diagram, depending only on its topology. The group acts on parameter space and the method determines the integral's dependence within group orbits. This paper analyzes the two-loop vacuum diagram. It is shown how the solution of the SFI equations practically  reproduces the most general value of the integral. On the way certain novel derivations are found, a geometrical interpretation is described, and divergences in general dimension are analyzed. These would hopefully be useful for engaging with more involved diagrams.}

\begin{document}

\maketitle

\section{Introduction}


The Symmetries of Feynman Integrals method \cite{SFI} considers a Feynman diagram of fixed topology, but varying masses, kinematical invariants and spacetime dimension. It associates to each diagram a set of differential equations in this parameter space. The equations define a Lie group $G$ naturally associated with the diagram, which acts on parameter space and foliates it into orbits. The diagram reduces to its value at some convenient base point within the same orbit plus a line integral over simpler diagrams (with one edge contracted). 
Given a diagram, the larger the group and the larger its orbits, the more useful the method is.

The SFI method is related to both the Integration By Parts method \cite{ChetyrkinTkachov1981} and the Differential Equations method \cite{DE}, see also the textbook \cite{SmirnovBooks}. The new elements include the definitions of the group and its orbits, as well as the reduction to a line integral. It allowed the determination of a novel 3-loop diagram with 3 mass scales \cite{VacSeagull}.

\begin{figure}
\centering \noindent
\includegraphics[width=10cm]{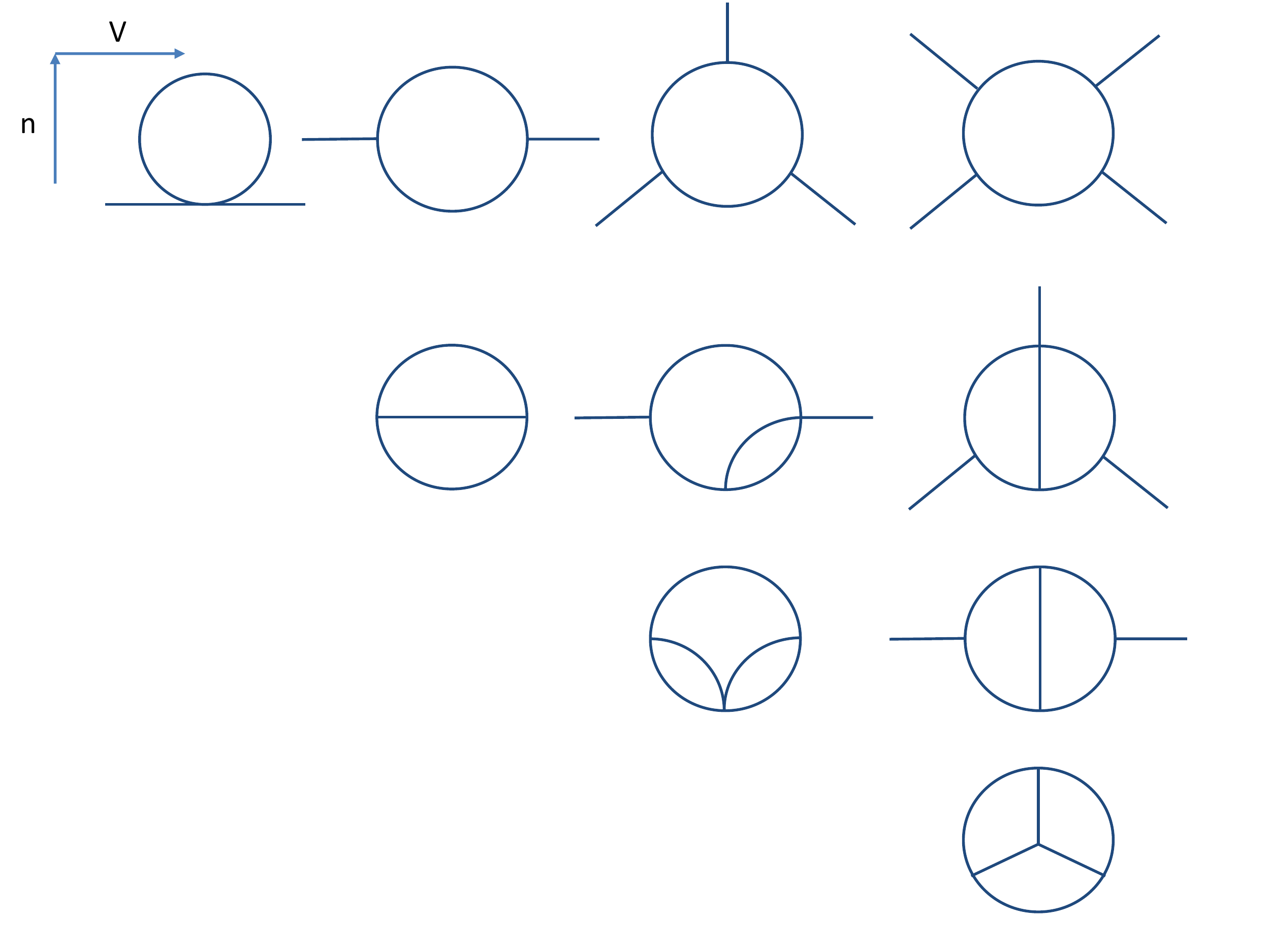} 
\caption[]{Hierarchy of diagrams according to SFI from left to right, namely according to ordering with respect to propagator contraction. 
Each column has diagrams of fixed number of vertices $V=1,2,3,4$. Since contraction reduces $V$ by unity the necessary sources for each diagram are always on its left. Each column in ordered according to the number of external legs $n$, which remains invariant under the contraction. 

Not all diagrams of given $V,n$ are shown. In particular each diagram can produce others with the same values of $V,n$ by adding propagators between existing vertices. For example the bubble diagram can produce in this way the sunrise and its generalizations.}
 \label{fig:DiagHierarchy} \end{figure}

In order to demonstrate the method and further develop it it is important to study specific diagrams. The inductive nature of the method where the value of a diagram is expressed in terms of simpler diagrams with one edge contracted suggests to (partially) order diagrams via this relation and to proceed step by step from simple to complex. Figure \ref{fig:DiagHierarchy} shows some of the simpler diagrams ordered by their inductive level. The tadpole on the leftmost of the figure is the simplest non-tree diagram and its evaluation is immediate through Schwinger parameters. The 1-loop propagator diagram, or the bubble, to its right can also be evaluated directly through the $\alpha$ variables. The SFI method was applied to it in \cite{bubble}. The group was found to be the 2*2 triangular real matrices and the orbit co-dimension was found to be 0 which allows to fully reproduce the diagram's value for general masses, kinematical invariant and spacetime dimension. The next diagram in this order is below it in the figure, namely the 2-loop vacuum diagram which we shall call the ``the diameter diagram'' and is the subject of this study.  The above-mentioned 3-loop diagram, the vacuum seagull \cite{VacSeagull}, is on the third column from the left at the bottom. And so on.

The diameter diagram was evaluated in full generality in \cite{FJJ1992} and \cite{DavydTausk1992} by using different methods and arriving at equivalent, yet different, expressions. Another equivalent expression was given at \cite{Davyd1999} inspired in part by the  geometrical interpretation of related diagrams \cite{DavydDelbourgo1997}. 
 The diameter is known to be closely related to the triangle diagram with massless propagators \cite{DavydychevTausk95magic}.
 
Why is it interesting to apply the SFI method to the diameter diagram? In this case the $G$-orbits have co-dimension 0 and therefore the method is expected to be maximally effective and could demonstrate itself, while at the same time providing tools for solving the SFI equations for other diagrams.

This paper is organized as follows. We start in section \ref{sec:setup} by setting up the problem and reviewing previous work within SFI including the SFI equation set. In section \ref{sec:mass_dep} the equations are solved in three different ways up to mass independent terms. In the following section these terms are determined through the consideration of base points thereby arriving at a full expression for the diagram's value (\ref{result}-\ref{def:contour}), which is equivalent to the known expressions. 
Inspired by \cite{DavydDelbourgo1997,Davyd1999} section \ref{sec:triangle} discusses the underlying geometry. Section \ref{sec:diverg} analyses this expression in terms of UV and IR divergences 
and on the way provides several additional tests for its correctness. Finally section \ref{sec:discussion} provides a summary of results and a discussion of the added value of the paper. 

\section{Setup}
\label{sec:setup}

\begin{figure}
  \centering \noindent
\includegraphics[width=4cm]{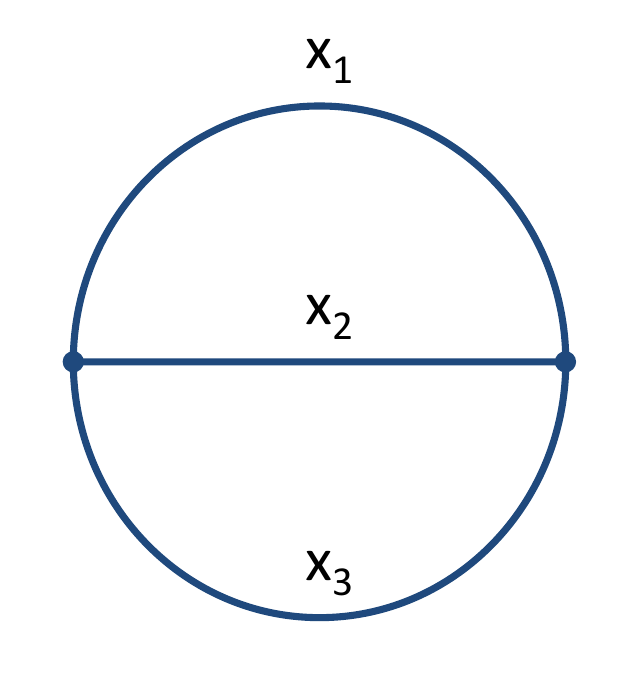} 
 \caption[]{The diameter diagram. ${x_1,x_2,x_3}$ denote the three mass-squared.}
  \label{fig:diam}
 \end{figure}

The two-loop vacuum diagram is shown in figure \ref{fig:diam}. For short we shall refer to it as the diameter diagram. The associated integral is defined by \be
 I\(x_1,x_2,x_3;d\) := \int \frac{d^d l_1\, d^d l_2}{\(l_1^2-x_1\) \(l_2^2-x_2\) \((l_1+l_2)^2-x_3\)} ~.
 \label{def:I}
 \ee
The parameter space is $X:=\(x_1,x_2,x_3\)$ consisting of the three possible masses-squared $x_i \equiv m_i^2$, and we consider a general spacetime dimension $d$. The discrete Feynman symmetry group induces a symmetry on parameters acting by $S_3$ permutations on the $x_i$ variables. 

This diagram has no potential numerators (namely, all quadratics in the two loop currents $l_1,\, l_2$ can be expressed as linear combinations of the three edge currents $k_1^2=l_1^2,\, k_2^2=l_2^2$ and $k_3^2=(l_1+l_2)^2$). Hence the SFI group $G$ saturates the group $GL(2,\IR)$ of general linear transformations among the loop currents \be
 G=GL(2,\IR) ~.
\label{group}
 \ee
The SFI equation set is given by  \bea
 0 &=& x_i \frac{\del}{\del x_i} I - (d-3) I \label{dimeq} \\
  0 &=& L_3\, I +(j_2-j_1) j_3'  \qquad  L_3:=x_1 \frac{\del}{\del x_1} - x_2 \frac{\del}{\del x_2}+(x_1-x_2) \frac{\del}{\del x_3}  \label{sl2eq} \label{def:L3} \\
 && \mbox{+2 equations gotten by cyclic permutations,} \nonumber
\eea
see \cite{SFI}, section 6. $L_i,\, I=1,2,3$ are differential operators. The $j$ functions which appear in the inhomogeneous part are defined through the value of the tadpole diagram as follows \bea
j_i  &:=& j(x_i) \non
j(x)  &:=& \raisebox{-14pt} {\includegraphics[width=2cm]{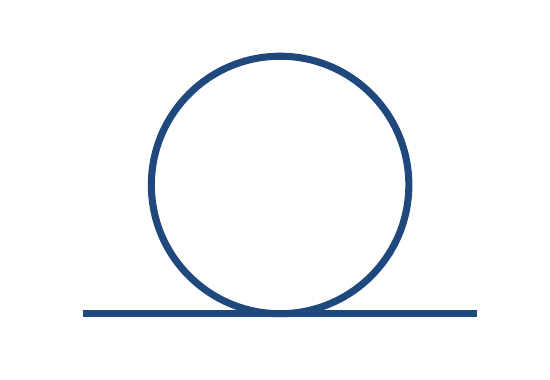}} = c_t\, x^a 
\label{def:j}
\eea
where
$a$ and $c_t$, the tadpole power and prefactor are defined by \bea
a &:=& \frac{d}{2} - 1 
\label{def:a} \\
c_t &:=& -i\, \pi^{d/2}\, \Gamma\(1-\frac{d}{2}\)
\label{def:ct}
\eea

The equation set consists of 4 equations corresponding to the generators of $gl(2,\IR) = \IR \bigoplus sl(2,\IR)$. The first equation (\ref{dimeq}) corresponds to the $\IR \simeq U(1)$ generator -- it is an instance of Euler's identity for homogeneous functions and is simply a consequence of dimensional analysis. The last three equations (\ref{sl2eq}) correspond to the $sl(2)$ generators.

In order to solve the equation set one should first identify the homogeneous solution $I_0$ and then apply the variation of the constants method. 
The equations corresponding to the $sl(2)$ generators are constant free (namely, $I$ appears in the equations only through its derivatives) as well as $d$ independent and hence the homogeneous solutions depends only on $sl(2)$ invariants. In fact, there is a single such invariant, given by the Heron formula or K\"all\'en invariant 
\be
\lambda := x_1^2+ x_2^2 + x_3^2 - 2 x_1 x_2 - 2 x_1 x_3 - 2 x_2 x_3
\label{def:lam}
\ee

The dimension equation (\ref{dimeq}) implies that the homogeneous solution is given by \cite{SFI} \be
I_0 =  |\lambda|^{\frac{d-3}{2}}
\label{def:I0}
\ee

Now the Feynman integral can be reduced to a line integral over the simpler diagrams (sources) through variation of the constants. The line integral is of the form
\be
I(x) = I_0(x) \int_{x_0}^x  \frac{S^\alpha(\xi)}{I_0(\xi)} d \xi_\alpha
\label{reduced}
\ee 
where $\xi_\alpha$ are coordinates on the $G$-orbit, and the $S^\alpha$ are the sources --  in the current case $\xi_\alpha=(x_1,x_2,x_3)$ and $S^\alpha$ are constructed from rational functions in $x$'s times the simpler diagrams $ j_i\, j'_k$.

At $\lam=0$ the equation set (\ref{dimeq},\ref{sl2eq}) degenerates into a algebraic equation. Hence this surface was called the algebraic locus and it allows to determine $I$ algebraically to be \cite{locus} \be
I|_{\lam=0} = \frac{\(x_2- x_3\) \(j_2 - j_3 \) j'_1 + \mbox{cyc.}}{(d-3) \( x_1 + x_2 + x_3 \)} 
\label{AlgLocSoln}
\ee
Physically, $\lam=0$ implies that $m_1=m_2+m_3$ or a permutation thereof and  it is a pseudo threshold of the Landau equations. The diagram's parameter space together with the algebraic locus is depicted in fig. \ref{fig:param_space}.

\section{The mass dependent part}

\label{sec:mass_dep}
 
In order to evaluate the Feynman integral for the diameter diagram $I$ defined in (\ref{def:I}), in this section we solve the equation set (\ref{dimeq},\ref{sl2eq}), and describe three alternative derivation paths. This leaves a freedom to add a homogeneous solution multiplied by a mass-independent function. In the following section we add into consideration certain base points which allow us to fully determine $I$.

We start by a minor revision of the equation set. Multiplying (\ref{sl2eq})  by $x_3$ one obtains
\be
x_3\, L_3\, I  = a\, c_t \( j_{13} - j_{23} \)
\label{sl2eq_v2}
\ee
where \be
j_{kl} := j(x_k x_l) = j_k\, j_l / c_t
\label{def:j_kl}
\ee
Two additional equations are gotten from (\ref{sl2eq_v2}) by cyclic permutations. This form has the advantage that each term on the right hand side (RHS) depends only on a single combination of variables, rather than two. 

Let us study the differential operator $L_3$ (\ref{def:L3}) and determine its invariants, namely functions $P=P(x_1,x_2,x_3)$ such that $L_3\, P = 0$. Since $X$, the parameter space is 3 dimensional, the space of invariants under single differential operator is generated by $3-1=2$ invariants, and since $L_3$ is homogenous in $X$ its invariants can be chosen to be homogenous. At linear order in $X$ $L_3$ has a single invariant  \be
 s^3 := (x_1 + x_2 - x_3)/2 
 \label{def:s} \ee
where the normalization factor was chosen for later convenience. Analogous quantities $s^1,\, s^2$ are defined by permutations. We shall give the $s^i$'s a geometric interpretation in section \ref{sec:triangle}.  The quadratic Heron / K\"all\'en invariant $\lam$ (\ref{def:lam}) is invariant by construction under all the $L_i$'s, namely \be
L_i\, \lam = 0 ~ i=1,2,3
\label{Llam}
 \ee
  so we choose it as the second invariant.

Having found the invariants $s^i,\, \lam$ we can express the quadratic variables on the RHS of (\ref{sl2eq_v2}) in terms of them, namely \be
x_1\, x_3 = (s^2)^2 - \lam/4
\label{x1x3s2}
\ee
and similarly for the other permutations. Moreover, the action of $L_i$ on $s^j$ is given by the expression \be
L_i\, s^j = -\epsilon_{ijk}\, x_k ~.
\label{Ls}
\ee

Now we can state and confirm the solution for the equations set (\ref{dimeq},\ref{sl2eq}). Derivations will be discussed later. The general solution is given by \be
I = c_d \[ B_1 + B_2 + B_3 + const\, I_0 \]
\label{soln1}
\ee
where \bea
 B_i &:=& B(s^i, \lam; \frac{d}{2}-1), ~~i=1,2,3 \non
 B(s,\lam;a) &:=& \int^s \( s'^2-\frac{\lam}{4} \)^{a-1} ds'
\label{def:B}
\eea
 and the diameter diagram prefactor is given by \be
  c_d := a \, c_t^{~2}  = \pi^d\,  \Gamma\(1-\frac{d}{2}\) \, \Gamma\(2-\frac{d}{2}\)
   \label{def:cd}
 \ee
where the $a, c_t$, the tadpole constants were defined in (\ref{def:a}-\ref{def:ct}).\footnote{
This $B$ is a dimensionful version of the one defined in \cite{bubble}, namely $B(s;a)=\int_0^s (1-s'^2)^{a-1} ds'$.}
 In the definition of $B$ the lower limit remains undefined, but can only be a function of $\lam$, which corresponds to the same freedom as adding a multiple of the homogenous solution $I_0$ (\ref{def:I0}).

It is straightforward to confirm that (\ref{soln1}) solves the equation set (\ref{dimeq},\ref{sl2eq_v2}) by using the identities (\ref{Llam},\ref{Ls}). It is the general solution since it contains an arbitrary multiple of the homogeneous solution.

It is more involved to derive this solution. In the following subsections we offer three derivations: the first is a straightforward solution through the method of characteristics; the second uses a symmetric ansatz; the third and final transforms the differential equations into (partially) invariant variables. Hopefully, various ingredients of these methods will be useful for a future analysis of more involved diagrams.

 
\subsection{Characteristics}

The first method uses the method of characteristics for solving linear first order partial differential equations (PDEs). The standard method is designed for a single PDE, while SFI defines an equation set and requires a generalization of the method. 

First we note that the three $sl(2)$ equations (\ref{sl2eq}) define characteristic surfaces given be constant values of the Heron/K\"all\'en invariant $\lam$ (\ref{def:lam}).  

Next we choose one of the $sl(2)$ generators and attempt to integrate its characteristic curves. This equation can be chosen such that the equation set for the curves decouple. We choose the generator  $L_3$. Its invariants are $\lam, s^3$ -- see (\ref{def:s}-\ref{Llam}). 

The associated equation set for the characteristic curves is given by \be
\frac{d}{dt} \[ \begin{array}{c}
x_1 \\
x_2 \\
x_3 \\
\end{array} \]
 = 
 \[ \begin{array} {c}
 x_1 \\
 -x_2 \\
 x_1 - x_2
 \end{array} \]  ~.
\ee
It decouples and its general solution is given by \bea
x_1(t) &=& x_{10}\, e^t \non
x_2(t) &=& \frac{(s^3)^2-\lam/4}{x_1(t)} \non
x_3(t) &=& x_1(t) + x_2(t) -2 s^3
\eea
where $x_{10}$ is an integration constant (and so are $\lam$ and $s^3$).

Next we solve for $I$ along the characteristic curve. Equation (\ref{sl2eq_v2}) has two source terms. We denote the solution for the first source by $c_d\, I_2$ (the diameter prefactor $c_d$ (\ref{def:cd}) was entered for later convenience and the notation $I_2$ will be justified later),  namely \be
L_3\,  c_d\,  I_2 = a\, c_t\, \frac{j_{13}}{x_3} 
\ee 
where the notation is defined in (\ref{def:L3}-\ref{def:ct},\ref{def:j_kl}). 
Along the curve \be
\frac{d}{dt} I_2 =  \frac{1}{c_t} \, \frac{j_{13}(t)}{x_3(t)}
\ee
and hence \be
I_2(t) =  \frac{1}{c_t}\,   \int^t  dt' \frac{j_{13}(t')}{x_3(t')} ~.
\label{def:I2}
\ee

We change parameterization (and hence the integration variable) according to \be
s^2 =(x_1(t) + x_3(t)-x_2(t))/2 = x_{10} e^t - s^3 ~.
\ee
The differential transforms into \be
dt = \frac{ds^2}{x_1(t)}
\ee
and the integral (\ref{def:I2}) becomes  \be
I_2 =  \int^{s^2} ds' \( s'^2 - \lam/4\)^{a-1} ~.
\label{res:I2}
\ee
Therefore $I_2 \equiv B_2$ defined in (\ref{def:B}), thereby justifying the notation $I_2$.  

Similarly the other source term defines $I_1$, namely $L_3 c_d I_1 = -a c_t j_{23}/x_3$, which is found to be a permutation of $I_2$ (\ref{res:I2}) . The remainder must be constant over the characteristic curves and therefore depends only on the $L_3$ invariants \be 
I_3=I_3(s^3,\lam) ~.
\label{I3_form}
\ee
 Altogether at this point we have \be
I = c_d \[ I_1(s^1,\lam) + I_2(s^2,\lam) + I_3(s^3,\lam) \]
\label{I_Ii}
\ee
where $I_3$ is an arbitrary function of its variables. However, now we may consider one of the other equations, say $L_1$, and in a similar way one finds that $I_3$ is also given by a permutation  of (\ref{def:I2},\ref{res:I2}) thereby completing the derivation of (\ref{soln1}-\ref{def:B}) through the method of characteristics. 

We note that, as it turns out, the same method of characteristics was used already by \cite{FJJ1992}, with a different choice of a generator. Specifically, that paper chooses the compact generator of $sl(2)$, namely $(x_2-x_3) \del_1 + (x_3-x_1) \del_2 + (x_1-x_2) \del_3 \equiv L_1 + L_2 + L_3$, and provides a well-written derivation. The paper also comments on another possible choice involving the dimension generator (\ref{dimeq}) -- see equation (4.2) there.  


\subsection{$S_3$ symmetry}

This derivation makes an essential use of the $S_3$ permutation symmetry of the diagram.

As in the method of characteristics we start with (\ref{sl2eq_v2}) and write the solution in the form (\ref{I_Ii}) where $I_1,\, I_2$ are particular solutions corresponding to the sources, namely \be
c_d\, x_3\, L_3 \[ \begin{array}{c} 
I_2 \\
I_1 \\
\end{array} \] = 
a\, c_t\, \[ \begin{array}{c} 
j_{13} \\
-j_{23} \\
\end{array} \] ~,
\label{IA_eq}
\ee
 and $I_3$ is a homogenous solution, namely \be
L_3\, I_3 = 0 ~.
\ee
 $I_3$ must be a function of $L_3$ invariants (\ref{I3_form}), namely of the form \be
I_3 = b(s^3,\lam)
\ee
where here we denote by $b(s^3,\lam)$ the yet undetermined function.

Notice that the source term $j_{13}$ is a function of $s^2, \lam$ only, namely \be
j_{13} = c_t \( (s^2)^2-\lam/4 \)^a
\ee

Guided by the $S^3$ symmetry it is reasonable to take $I_2$  to be a permutation of $I_3$, and guided by the form of the source, it is taken to be a function of $s^2,\lam$ namely  \be
I_2 = b(s^2,\lam) ~.
\ee
thereby justifying the notation $I_2$.

Substituting into the first component of (\ref{IA_eq}) the LHS becomes \be
c_d\, x_3\, L_3\, b = c_d\, x_3\, (L_3 s^2) \frac{\del}{\del s^2} b = + c_d\, x_3\, x_1 \frac{\del}{\del s^2} b
\ee
where in the last equality we used the identity (\ref{Ls}).  The RHS can be written as \be
a\, c_t\, j_{13} = c_d \( (s^2)^2-\lam/4 \)^a
\ee

Equating LHS and RHS and recalling (\ref{x1x3s2}) we find that a solution is possible and is given by \be
b(s^2,\lam) = \int^{s^2} ds' \( s'^2-\lam/4 \)^{a-1} \equiv B(s^2,\lam)
\ee
namely, the function $b(s^2,\lam)$ is the same as the function $B(s^2,\lam)$ defined at (\ref{def:B}). 

Similarly \be 
I_1 = B(s^1,\lam) \
\ee
solves the second component of (\ref{IA_eq}). 

Summarizing, we have arrived at the permutation symmetric form (\ref{soln1}-\ref{def:B}) and have provided another derivation of it.

\subsection{Partial invariants}

The idea here is to use coordinates which are compatible with the group, namely group invariants. In our case we would like to use $\lam$ the $SL(2)$ invariant defined in (\ref{def:lam}). By definition $\del/\del \lam$ cannot appear in the 3 $sl(2)$ equations. It order to complete $\lam$ into a set of 3 coordinates on $X$ we must break the  diagram's $S_3$  symmetry. Inspired by a similar change of variables for the bubble diagram \cite{bubble} we start by changing \bea
 (x_1,\, x_2,\, x_3)  \to (\Delta:=x_1-x_2,\, x_3,\, \lam)
 \label{coor1}
 \eea
These coordinates have the advantage that the inverse transformation does not involve square roots, and is given by \bea
x_1 &=& \frac{\(x_3 + \Delta \)^2-\lam}{4 x_3} \non
x_2 &=& \frac{\(x_3 - \Delta \)^2-\lam}{4 x_3}
\eea

In these coordinates the $sl(2)$ equation set (\ref{sl2eq}) translated to \be
\[ \begin{array}{c}  
 \frac{\del}{\del \Delta} \\
  \frac{\del}{\del x_3}
\end{array} \]
 I = \frac{c_t}{2\, x_3} \[ \begin{array}{c}
 x_3\, j'_{13} - x_3\, j'_{23} + \Delta\, j'_{12} \\
 x_3\, j'_{13} + x_3\, j'_{23} - (x_1 + x_2) j'_{12}
 \end{array} \]
 \label{sl2eq2}
 \ee
 where $c_t$ is defined in (\ref{def:ct}) and $j_{kl}$ in (\ref{def:j_kl}). The dimension equation (\ref{dimeq}) becomes \be
 \[ 2 \lam \frac{\del}{\del \lam} + \Delta \frac{\del}{\del \Delta} + x_3 \frac{\del}{\del x_3} \]I = (d-3) I 
 \label{dimeq2}
 \ee
 
These equations in the $\lam,\, \Delta,\, x_3$ variables can now be integrated. However, we shall choose to perform another redefinition of variables in order to gain some further simplification, and hence here it suffices to present an outline of the integration in $(\lam,\, \Delta,\, x_3)$. One starts with integrating the equation for $\del I/\del \Delta$, the top component of (\ref{sl2eq2}). The arguments of the $j'_{13}$ function on the RHS is written as $x_1 x_3 =  \((x_3 + \Delta)^2-\lam\)/4$ and then the first term on the RHS integrates to a function of the form $\int Q(\Delta)^{\frac{d}{2}-1} d\Delta$ where  $Q(\Delta)$ is some quadratic function. The second term is analogous after exchanging $x_1 \leftrightarrow x_2$, while the third term appears different, but becomes similar after changing the integration variable to be $\Delta^2$. In fact, the three terms of $I$ computed in this way transform like $x_i$ under $S_3$. This is a consequence of the $S_3$ symmetry, though it appears surprising in the $(\Delta,x_3,\lam)$ variables. The integral is unique up to a constant which could depend on $x_3$ and $\lam$ yet the second component of (\ref{sl2eq2}) implies that it is independent of $x_3$ and finally the dimension equation (\ref{dimeq2}) implies that it is proportional to $\lam^{(d-3)/2}$, namely it is proportional to the homogeneous solution (\ref{def:I0}), which is the expected residual freedom.

Motivated by the form of the source we now proceed to redefine \be
(\Delta,\, x_3,\, \lam)  \to (s^1,\, s^2,\, \lam)
\ee
through (\ref{def:s}) and its permutations, namely  \be
s^{1,2} = (\pm \Delta + x_3)/2 ~.
\ee
  
In these variables the $sl(2)$ equations become \be
\[ \begin{array}{c}  
 \frac{\del}{\del s^1} \\
  \frac{\del}{\del s^2}
\end{array} \]
 I = c_t \[ \begin{array}{c}
  j'_{23} - \frac{(s^2)^2-\lam/4}{(s^1+s^2)^2}\, j'_{12}\(s^1,s^2,\lam\) \\
  j'_{13} - \frac{(s^1)^2-\lam/4}{(s^1+s^2)^2}\, j'_{12}
 \end{array} \]
 \label{sl2eq3}
 \ee

The first source term in each equation integrates directly implying \be
 I = c_d \[  B(s^1,\lam;\frac{d}{2}-1) + B(s^2,\lam;\frac{d}{2}-1) + \tilde{I} \(s^1,s^2\) \]
 \ee
 where $B(s,\lam,a)$ is defined in (\ref{def:B}) and $c_d$ in (\ref{def:cd}) and where finally the remaining unknown $\tilde{I}$ satisfies \be
 \[ \begin{array}{c}  
 \frac{\del}{\del s^1} \\
  \frac{\del}{\del s^2}
\end{array} \]
 \tilde{I} = - \[ \begin{array}{c}
  \frac{(s^2)^2-\lam/4}{(s^1+s^2)^2}\,  \\
 \frac{(s^1)^2-\lam/4}{(s^1+s^2)^2}
 \end{array} \] j'_{12} ~.
 \label{tilde-I-eqs}
 \ee  
 
 The $S_3$ permutation symmetry suggests that \be
 \tilde{I}=B(s^3,\lam;\frac{d}{2}-1)
 \label{def:tilde-I}
 \ee
 where $s^3$ is considered as a function of $s^1, s^2, \lam$ namely \be
 s^3:= - \frac{\lam/4 + s^1 s^2}{s^1 + s^2} ~.
 \ee
 Indeed the partial derivative \be
 \(\frac{\del s^3}{ \del s^1} \)_{s^2,\lam} = - \frac{(s^2)^2-\lam/4}{(s^1 + s^2)^2}
 \ee
 and similarly with $1 \leftrightarrow 2$ guarantees that (\ref{def:tilde-I}) satisfies (\ref{tilde-I-eqs}).
 
Summarizing the general solution to the SFI equations (\ref{dimeq}-\ref{sl2eq}) is \be
I = c_d \[  B(s^1) + B(s^2) + B(s^3) \] 
\ee
where $B(s,\lam;a=d/2-1)$ is defined in (\ref{def:B}) and the diameter prefactor is defined in (\ref{def:cd}). In this expression the lower limit in the definition of $B$ remains undefined which corresponds to the freedom to add a multiple of the homogenous solution (\ref{def:I0}).

\section{The mass universal part}
\label{subsec:mass_universal}

In the previous section the general solution for the SFI equations was obtained (\ref{soln1}). Since it contains an arbitrary multiple of the homogeneous solution, the SFI equations must be supplemented by some boundary conditions in order to fully evaluate the integral (\ref{def:I}). 

Geometrically, the action of the SFI group $G$ on the parameter space $X$ foliates it into $G$-orbits, namely orbits of the group where each point on the orbit can be transformed to any other point on it through the group action. For the diameter diagram the group is \be
G = GL(2,\IR) \simeq SL(2,\IR) \times \IR ~.
\label{def:G}
\ee
The $SL(2,\IR)$ subgroup preserves the Heron/K\"all\'en invariant $\lambda$ defined in (\ref{def:lam}), which is a quadratic form with signature (2,1). Hence we may identify $SL(2,\IR) \equiv SO(2,1)$, namely the $SL(2)$ factor acts on $X$ space through Lorentz transformations. Therefore the $SL(2)$ orbits are surfaces of constant $\lam$. The $\IR$ factor acts as radial rescaling (this is true for all diagrams).

\begin{figure}
\centering \noindent
\includegraphics[width=8cm]{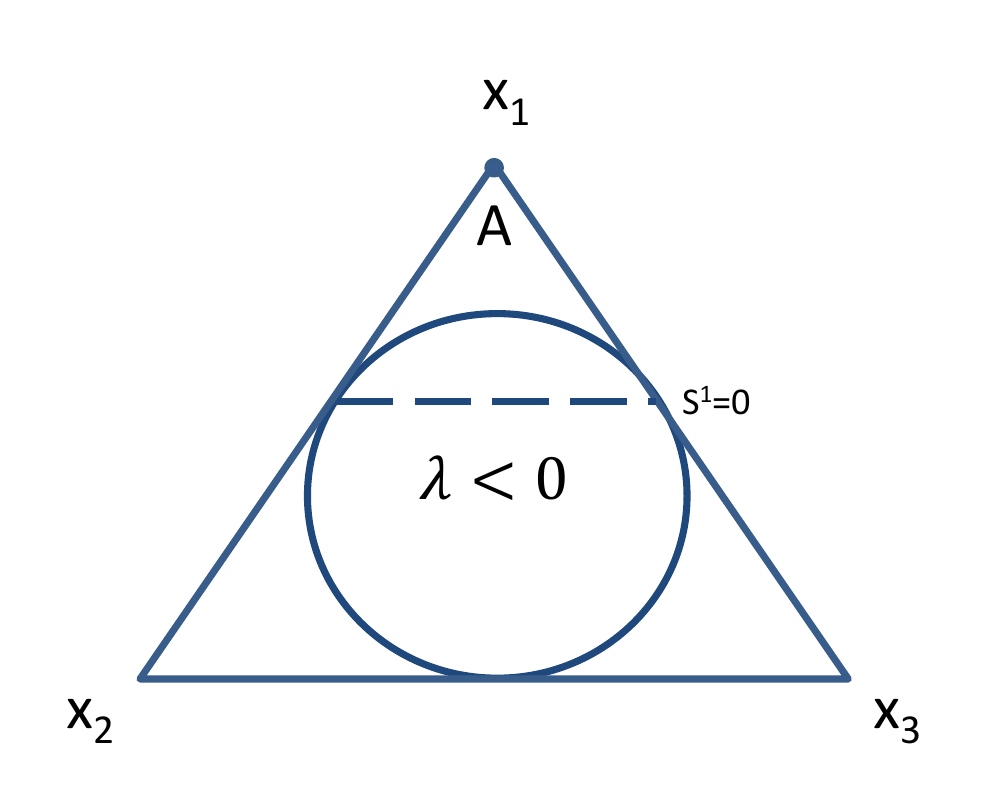} 
\caption[]{The parameter space for the diameter diagram. More precisely it is the projective version gotten by taking the section through the plane $x_1+x_2+x_3=const$. The triangle vertices represent the axes, for example the top vertex (denoted by A) represents the $x_1$ axis and the bottom edge represents the $x_1=0$ plane. The circle represents the $\lam=0$ cone. The dashed line denotes the $s^1=0$ plane, where $s^i$ are defined in (\ref{def:s}).}
\label{fig:param_space}
\end{figure}

Fig. \ref{fig:param_space} summarizes the foliation of parameter space into G-orbits. Restricting our attention to the quadrant of positive mass-squared $x_1, x_2, x_3 \ge 0$ represented by the inside of the triangle in the figure,\footnote{
This is a natural condition and it keeps the $j$ sources away from their branch cut, and so single valued.} 
we note the $\lam=0$ cone which is represented by a circle in the figure. This cone, which is a light cone of the above mentioned $2+1$ Minkowski space is a degenerate 2d orbit of $G$. It separates the positive quadrant into the following 3d orbits \bi
\item Outside the cone, namely positive $\lam$ (``spacelike''). This region is separated by the quadrant into three parts which are related to each other through the $S_3$ permutation symmetry.
\item Inside the cone, namely negative $\lam$ (``timelike'').
\ei 

The SFI equation set requires a boundary condition in order to determine for each orbit the free constant in the general solution (\ref{soln1}), as we now turn to discuss.

\subsubsection*{Outside the cone}

For concreteness we shall consider the upper region between the circle and the triangle in the figure, where $m_1>m_2+m_3$.

In this region we choose as base point the (projective) point $m_2=m_3=0$, denoted by $A$ in the figure. At this point the diagram can be readily evaluated using direct integration in the Schwinger parameter space (see appendix \ref{app:eval_A}) to \be
I_A = \pi^d\, \mu^{d-3}\, \Gamma\(2-d\)\, \frac{2\pi}{\sin  \frac{d}{2}\pi} 
\label{I_A}
\ee  
where the parameter $\mu$ defines all the masses as follows \bea
x_1 &=& \mu \non
x_2 = x_3 &=& 0
\label{A_coord}
\eea

\begin{figure}
\centering \noindent
\includegraphics[width=8cm]{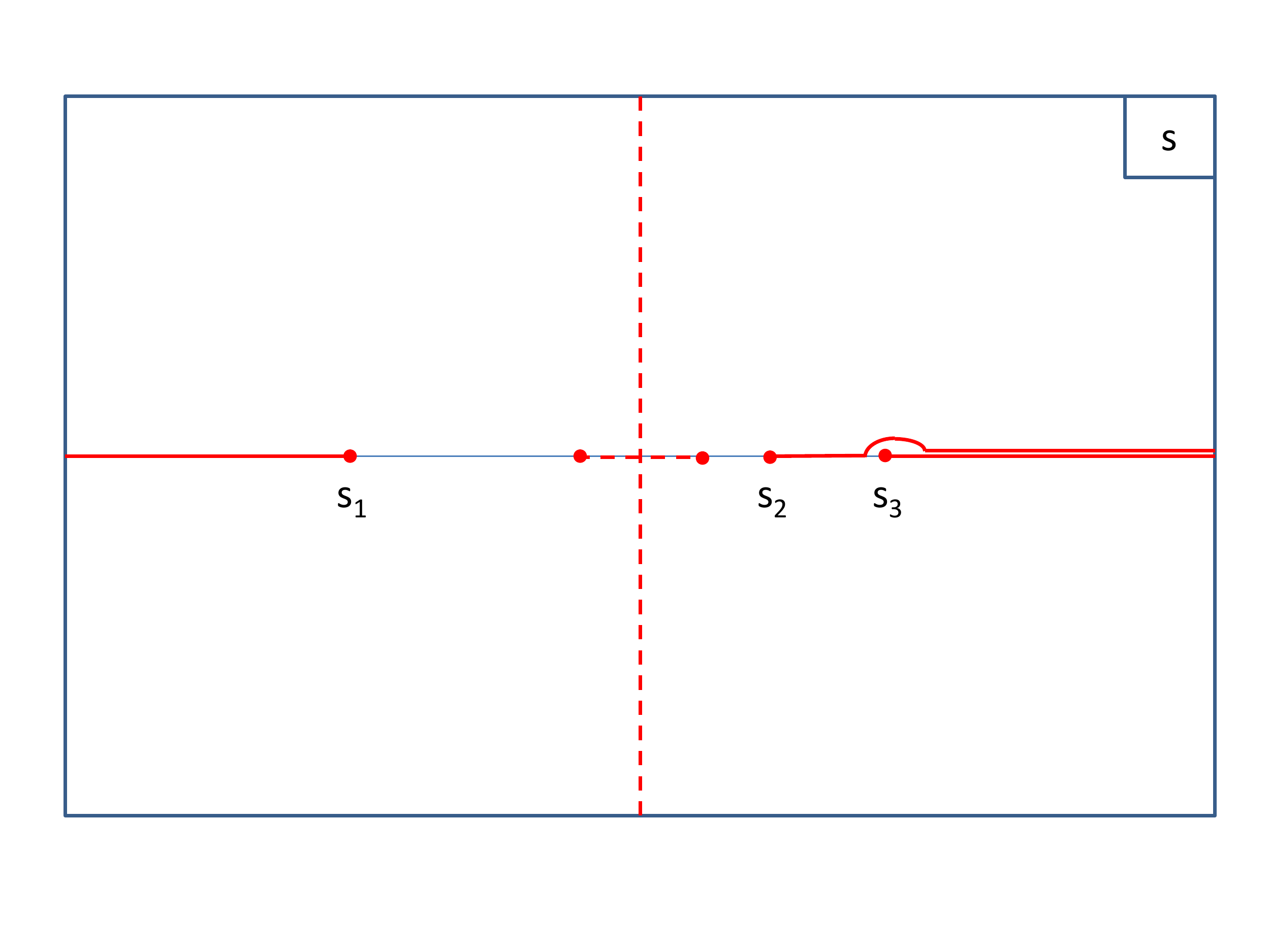} 
\caption[]{Contours of integration in the $s$ plane for the upper region outside the cone in $X$ space, namely positive $\lam$ and $m_1>m_2+ m_3$. The branch cuts are shown by dashed lines.}
\label{fig:s_plane_1}
\end{figure}

In order to compare with the general solution (\ref{soln1}) we must specify the so-far unspecified contours of integration in the complex $s$ plane. The integrand $\(s^2- \lam/4\)^{a-1}$ has branch cuts which we choose conventionally to be where $s^2-\lam/4$ is real and negative. For positive $\lam$ this determines the branch cuts in the $s$ plane to be cross-shaped as shown in figure \ref{fig:s_plane_1}, namely either $\Re(s)=0$ or $\Im(s)=0$ and $\left| \Re(s) \right| \le \sqrt{\lam}/2$. 

Next we wish to locate $s^1, s^2, s^3$ relative to the branch cuts. $s^1=0$ determines a plane in parameter space shown as a dashed line in fig. \ref{fig:param_space} (it passes through the points $x_1=x_2, x_3=0$ and $x_1=x_3, x_2=0$ which are the points where the circle is tangent to the triangle). Therefore within the region under discussion we have negative $s^1$ and positive $s^2, s^3$ (more precisely $s^1 \le -\sqrt{\lam}/2, s^2,s^3 \ge \sqrt{\lam}/2$). 

Each $s^i$ is the end-point of a contour. The starting points are arbitrary and their choice fixes the coefficient of the homogeneous solution. On dimensional grounds the starting point can only be a multiple of $\sqrt{\lam}$. For simplicity and with hindsight we choose all starting points to be at infinity $-\infty$ for $s^1$ and $+\infty$ for $s^2, s^3$, as shown in the figure, so that the contours do not cross the branch cut.

In summary, we have chosen the contour $C$ in the $s$ plane to be a formal sum of three segments as follows \bea
\int_C &:&= \int_{-\infty}^{s^1} + \int_{+\infty}^{s^2}  + \int_{+\infty}^{s^3} = \non
	&=& \int_{-\infty}^{s^1} -  \int_{s^2}^{+\infty} - \int_{s^3}^{+\infty}
\label{def:cont}
\eea

In order to determine the coefficient of the homogenous solution we evaluate the integral at point A (\ref{A_coord}). The values of the $s^i$ variables are $s^2=s^3=-s^1=\mu/2$ and $\lam$ is given by $\lam=\mu^2$. Integrating \bea
 \tilde{I}_A &:=& \int_C \(s'^2-\lam/4 \)^{a-1} ds' = - \int_{\mu/2}^{+\infty}  \(s'^2-\mu^2/4 \)^{a-1} ds' = \non
  		&=&  - \( \frac{\mu}{2} \)^{2a-1} \int_1^\infty \( \tilde{s}^2-1\) d\tilde{s}  = \non
	&=&  - \( \frac{\mu}{2} \)^{2a-1}\, \half B\(a,\half-a\)
 \label{I-tilde_A}
  \eea 
where $B(a,b)$ denotes the Beta function and we changed variables $s \to \tilde{s} := 2 s/\mu$ to reach the second line and then $\tilde{s} \to t:=1/\tilde{s}^2$ in order to reach the third.

Combining (\ref{soln1}, \ref{def:cd}, \ref{I_A},\ref{I-tilde_A}) we find \be
c_d\, const\, I_0 = c_d\, \tilde{I}_A - I_A = 0
\ee
and hence $const=0$. 

Summarizing, in this region \be
I = c_d \[ \int_{-\infty}^{s^1} -  \int_{s^2}^{+\infty} - \int_{s^3}^{+\infty} \] \(s'^2-\lam/4 \)^{a-1} ds'
\ee
where $a,\, s^i,\, c_d,$ were defined in (\ref{def:a}, \ref{def:s}, \ref{def:cd}), respectively. 

\subsubsection*{Inside the cone}

\begin{figure}
\centering \noindent
\includegraphics[width=8cm]{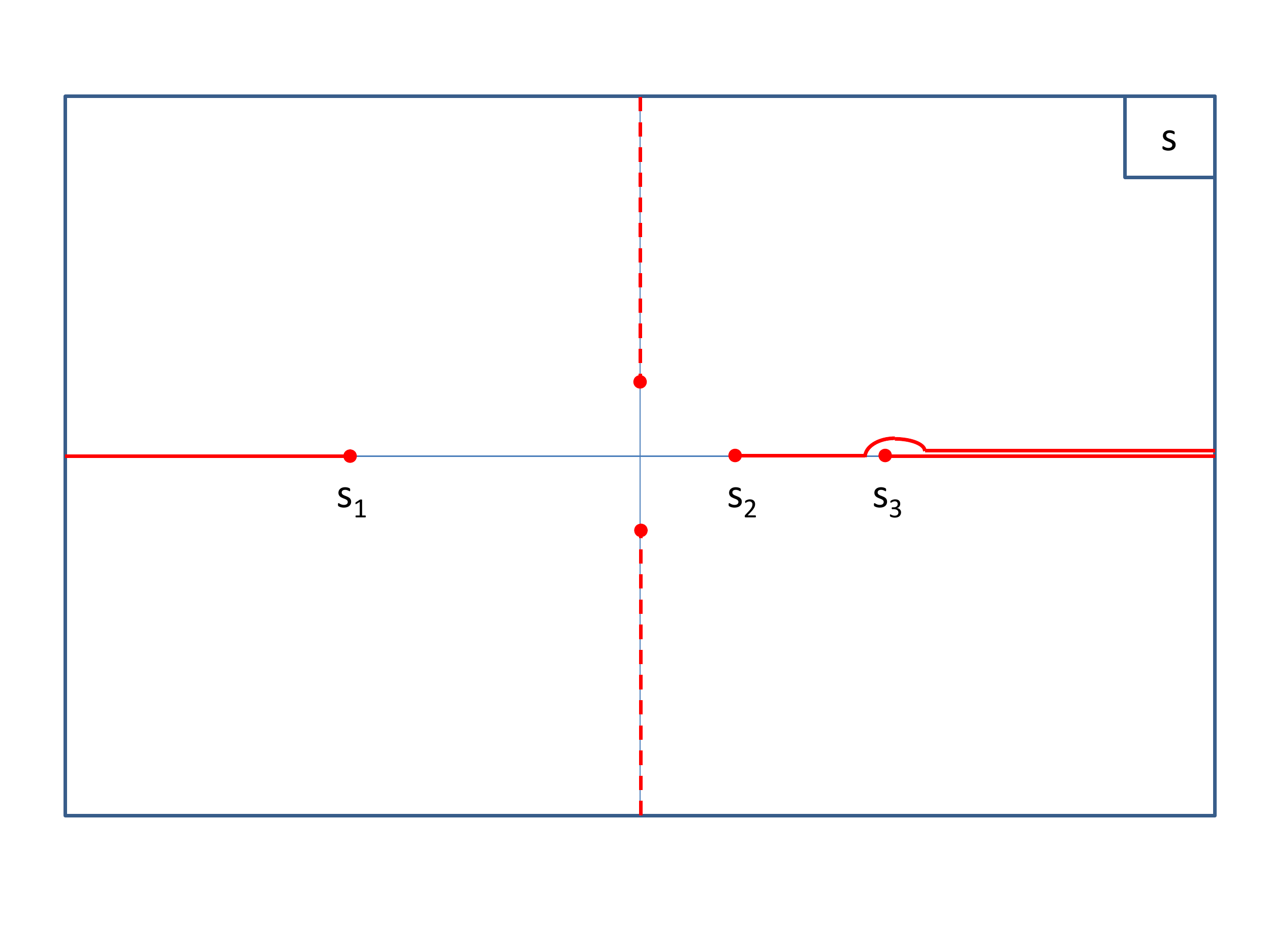} 
\caption[]{Contours of integration in the $s$ plane for the inside of the cone in $X$ space, namely negative $\lam$. 
The branch cuts shown by dashed lines do not intersect the real $s$ axis anymore.}
\label{fig:s_plane_2}
\end{figure}

We turn to the region inside the cone, shown as the inside of the circle in fig. \ref{fig:param_space}. Now $\lam$ is negative and the cut structure in $s$ space, which is determined by the locus of negative values of $s^2-\lam/4$, opens up to become fig. \ref{fig:s_plane_2} where the branch cut extends along $\Re(s)=0$ for $\left| \Im(s) \right| \ge \sqrt{-\lam}$ (compare with fig. \ref{fig:s_plane_1}). 

Now there are no cuts on the real $s$ axis and so the contour $C$ remains well-defined. However, (\ref{def:cont}) is not symmetric in the $s^i$. That was justified outside the cone since the variables differed in their signs, but the inside of the cone is symmetric. 

Still, it is possible to rewrite the analytic continuation of $C$ in a manifestly $S_3$ symmetric way, namely \bea
\int_C &=& \int_{-\infty}^{s^1} -  \int_{s^2}^{+\infty} - \int_{s^3}^{+\infty} = \non
	&=& \int_{-\infty}^{+\infty} -  \int_{s^1}^{+\infty} - \int_{s^2}^{+\infty} - \int_{s^3}^{+\infty} 
\label{def:cont2}
\eea
where now the first segment which runs all over the real $s$ axis does not cross a cut and hence is well defined. 

This contour is a viable candidate for setting the constant, namely the mass universal part in the region inside the cone. However, this cannot be inferred from the differential equations alone, since they relate only values lying on the same $G$ orbit, while by construction the inside and the outside of the cone belong to different orbits. 

We determine the constant by comparison with expressions in the literature. Comparison with \cite{FJJ1992} eq. (4.16) shows that (\ref{def:cont2}) is indeed valid, see appendix \ref{app:FJJ_inside} for details. In addition, we found agreement with the results of \cite{DavydTausk1992} obtained through the Mellin-Barnes transform, by comparing at the equal mass base-point $x_1=x_2=x_3$ (the center of the circle in fig. \ref{fig:param_space}), see e.g. eq. (21) of \cite{Davyd1999}.

We note that the constant was set through comparison with results outside the SFI method. I expect that the SFI method can be supplemented by a boundary condition for the SFI differential equations at the singular locus which would allow an alternate derivation. In fact, the current result may be helpful towards formulating such a boundary condition.

\subsection{The full expression}

Summarizing all the preceding derivations, the most general diameter integral defined in (\ref{def:I}) is given by \be
I(x_1,x_2,x_3;\, d) = c_d  \int_C \(s'^2-\lam/4 \)^{d/2-2} ds' 
\label{result}
\ee
where the diameter constant $c_d$ is given by (\ref{def:cd}) and the contour $C$ depends on the region in parameter space and is given by 
 \be 
\int_C =\begin{cases}
\int_{-\infty}^{s^1} -  \int_{s^2}^{+\infty} - \int_{s^3}^{+\infty}		& ~~~\mbox{Outside the cone where } \lam>0  \\
												& ~~~\mbox{ and } m_1>m_2+m_3 \\
\int_{-\infty}^{+\infty} -  \sum_{i=1}^3 \int_{s^i}^{+\infty}		&  ~~~\mbox{Inside the cone where } \lam<0 \\
\end{cases} ~.
\label{def:contour}
\ee
The $s^i$ variables are defined in (\ref{def:s}). Inside the cone, where both contour definitions avoid branch cuts, they coincide.

Note that (\ref{result}) can be written as \be
I = c_t \int_C j'\(s'^2-\lam/4\)\, ds'
\ee
where $j$ is the bubble diagram (\ref{def:j}), thereby stressing that (\ref{result}) represents a line integral over simpler diagrams, just like SFI leads us to expect.
 
For completeness we recall that the integral \be
 \int_s^\infty  \(s'^2-\lam/4 \)^{d/2-2} ds' 
 \label{quad_int}
\ee
 can be expressed by the incomplete Beta function $B\(x;a,b\)$, as follows. For $\lam \ge 0$  (\ref{quad_int}) is given by the following two alternative expressions \bea
&&  \half \(\frac{\lam}{4}\)^{\frac{d-3}{2}} \[ B\( \frac{d}{2}-1,\frac{3-d}{2}\) - B\( 1-\frac{\lam}{4s^2} ; \frac{d}{2}-1,\frac{3-d}{2}\) \]   = \non
 &=&  \lam^{\frac{d-3}{2}} \[ B\( \frac{d}{2}-1,3-d\) - B\( 1-\frac{\sqrt{\lam}}{s+\sqrt{\lam/4}} ; \frac{d}{2}-1,3-d\) \]
 \eea
 where $B\(a,b\):=B\(1;a,b\)$ is the (complete) Beta function. We note that at point A (\ref{A_coord}) all the incomplete Beta function vanish and $I$ is given by a complete Beta function (\ref{I-tilde_A}). For $\lam \le 0$ (\ref{quad_int}) is given by \be
  \half \(\frac{|\lam |}{4}\)^{\frac{d-3}{2}} \[ B\( \half,\frac{3-d}{2}\) - B\( 1-\frac{|\lam |}{|\lam |+4s^2} ;\half,\frac{3-d}{2}\)\] 
  \ee
 The incomplete beta function, in turn,  is a special case of the hypergeometric function ${}_2 F_1$ \be
 B(x; a,b) = \frac{x^a}{a}\,  {}_2 F_1 (a,1-b;\, a+1;\, x) ~.
 \ee
  For more properties of these functions see \cite{bubble} eq.s (4.15-4.21).

\section{Underlying triangle geometry}
\label{sec:triangle}

\begin{figure}
\centering \noindent
\includegraphics[width=6cm]{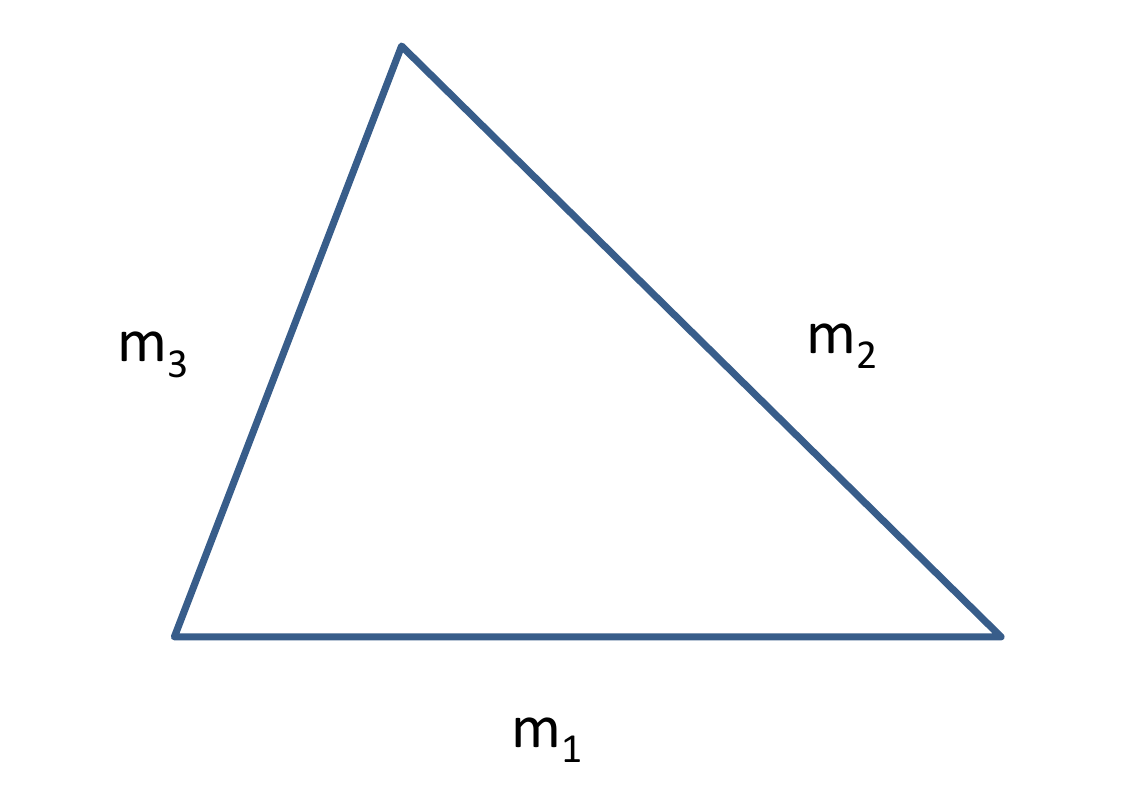} 
\caption[]{A triangle whose sides are $m_1, m_2, m_3$ is the geometry underlying the diameter diagram. See text for further details.}
\label{fig:triangle}
\end{figure}

\cite{DavydDelbourgo1997} introduced a geometric approach to Feynman integrals linking them to certain polyhedra, see also \cite{Davyd2016}, and \cite{Davyd1999} related it to the diameter diagram. Following them we interpret the end expression (\ref{result}) in terms of the geometry of a triangle whose sides are $m_1, m_2$ and $m_3$ as shown in fig. \ref{fig:triangle}.

The case of the diameter diagram is very close to the 1-loop propagator diagram (``bubble'') which was discussed in \cite{DavydDelbourgo1997}, see also \cite{bubble} for a treatment through SFI where the underlying triangle geometry is discussed in section 5. 

When $m_1, m_2, m_3$ satisfy the triangle inequalities they define a unique triangle in Euclidean plane (in fact, since each edge is timelike, we should think of it as a 2-time plane). Moreover, when the triangle inequality is violated, for example if $m_1>m_2+m_3$ a unique triangle is defined in a $1+1$ Lorentzian space \cite{bubble}. The first case occurs inside the $\lam=0$ cone in parameter space, while the second occurs outside. Altogether any point in $X$ determines a unique triangle in a plane of a determined signature.

The quantity $\lam$ is related to the area $A$ of the (Euclidean) triangle  through Heron's formula, namely \be
A^2 = -\frac{1}{16} \lam ~.
\ee

The quantity $s^1:=(x_2+x_3-x_1)/2$ is interpreted after recalling that $x_i \equiv (m_i)^2$ and using the law of cosines to be \be
 s^1 = m_2 m_3\,  \cos \alpha_1
 \ee
 where $\alpha_1$ is the angle between $m_2$ and $m_3$ . In addition, $s^i$ is proportional to the Legendre conjugate of $x_i$ with respect to  $\lam$, namely \be
 s^i = -\frac{1}{4} \frac{\del}{\del x_i} \lam 
\ee 
and the Legendre transform is \be
\lam=-4 \( s^1\, s^2 + s^1\, s^3 + s^2\, s^3 \)
\label{lam_s}
\ee 
 Finally we recall that by construction \be 
 (s^1)^2-\lam/4 = (s^1+s^2)(s^1+s^3) = x_2 x_3 ~,
 \ee
where the Left Hand Side is related to the integrand of the result.


\section{Divergences}
\label{sec:diverg}

This section analyses the expression for the value of the diagram (\ref{result}) focussing on divergences in various dimensions. Part of the motivation is to provide independent tests for the expression beyond the comparison with the literature.

Dimensional regularization is known to translate divergences into poles in the $d$ plane. The current approach which strives to determine the dependence of the integral on all of its possible parameters, allows one to clearly distinguish between different types of divergences. \emph{UV divergences} are independent of masses and kinematical invariants and hence \emph{are generic in parameter space}. Conversely, IR divergences are local in parameter space, namely they occur when some condition is met.

\presub {\bf Lemma}. Many of the pole series below will be derived through a uniform mechanism, which we turn to describe. Consider an integral of the form \be
\int^\infty dy\, y^{a(d)}\, \[ 1 + {\cal O} \(\frac{\mu}{y^c} \) \]
\ee
where $y$ is any integration parameter, $d$ is the spacetime dimension, $a(d)$ is a function of $d$ which is normally linear in $d$, namely $a(d)=a\, d+ b$, $\mu$ is a parameter and finally $\mu/y^c,\, c>0$ is the expansion parameter of the integrand near $y=\infty$. A pole appears at order $k$ of the expansion when \be
1 + a(d_k) - k\, c = 0 \qquad k=0,1,2,\dots
\ee
namely \be
d_k = \frac{1}{a} \( -b-1 + k\, c \)   ~. 
\label{poles_infty}
\ee

A closely related case is an integral near $0$ of the form \be
\int_0 \frac{dy}{y^{a(d)}}\, \[ 1 + {\cal O} \(\frac{y^c}{\mu} \) \] ~.
\ee
This time poles appear at \be
1 - a(d_k) + k\, c = 0 \qquad k=0,1,2,\dots
\label{poles_zero}
\ee

\subsection*{UV divergences}

Here we shall study the UV divergences first from the definition of the integral (\ref{def:I}) and the associated parameter representation (Schwinger parameters), and then from the expression for its value (\ref{result}).

{\bf $\mathbf{l,\al}$ plane}. Consider (\ref{def:I}) and the associated parametric representation \bea
 I &=& \int_0^\infty d\al^1 d\al^2 d\al^3\, U\( \{\al^i\}_{i=1}^3 \)^{-d/2} \non
 U &:=& \al^1 \al^2 + \al^1 \al^3 + \al^2 \al^3
\label{def:I_param}
\eea 
 where $U$ is the Kirchhoff-Symanzik polynomial. 
 
UV divergences can arise from several regions of integration. The first region is $l_1, l_2 \to \infty$ where the integral behaves as \be
 I \sim \int^\infty \frac{ d(l^2)\, (l^2)^{d/2-1}}{(l^2)^3} \[1 + {\cal O} \( \frac{\mu}{l^2} \) \] ~,
 \ee
 and therefore (\ref{poles_infty}) implies a series of poles at $d=3,4,5,\dots$. Equivalently,  the corresponding region in $\alpha$ parameter space is $\al^1,\al^2,\al^3 \to 0$. Defining $\al=\al^1 +\al^2+\al^3$ and rescaling to define $\beta^i$ through $\al^i = \al \cdot \beta^i$  the integral (\ref{def:I_param}) is found to behave as \be
 I \sim \int_0 (\al)^2 d\al\, (\al)^{-d} \exp (-\mu \al) \times \dots  
 \label{I_al_rad}
 \ee
where the ellipsis denote an integration over the $\beta^i$ variables. Using (\ref{poles_zero}) we reproduce the above mentioned list of poles.  

The second region is $l_1 \to \infty, l_2 \mbox{ fixed}$. An analogous procedure yields the pole series $d=4,6,8,\dots$. Due to the multiplicative structure of (\ref{I_al_rad}) the two series add in multiplicity to produce $d=3,4^2,5,6^2,\dots$ where throughout this section superscripts denote pole multiplicity. Other single loop regions such as  $l_2 \to \infty, l_1 \mbox{ fixed}$, do not provide additional poles.

The analysis of the pole structure in the $l / \al$ plane is summarized in the following table \be 
\begin{array}{ccc}
\mbox{Region } & \mbox{pole series } & \mbox{comments } \\
 \hline 
 l_1, l_2 \to \infty 	& 3,4,5,\dots 	& \mbox{ equiv. to } \al^1,\al^2,\al^3 \to 0 \\
  l_1 \to \infty, l_2 \mbox{ fixed} 	& 4,6,8,\dots & \mbox{ equiv. to } \al^1,\al^3 \to 0,\, \al^2 \mbox{ fixed} \\
\\ \hline 
\mbox{Total } & 3,\, 4^2, 5,\, 6^2, \dots & \\
\hline
\end{array}
\label{tab:UV1}
\ee

{\bf $\mathbf{s}$ plane}. The expression (\ref{result}) is a product of $c_d$, the diameter constant, with a function which depends on $s^1,s^2,\, s^3$. The poles of $c_d$ (\ref{def:cd}) are recognized to be $d=2, 4^2, 6^2, 8^2,\dots$.

 The $s$ integral behaves as \be
I \sim \int^\infty  \(s'^2-\lam/4 \)^{d/2-2} ds' \sim \int^\infty ds'\, s'^{d-4} \[1 + {\cal O} \( \frac{\lam}{s^2} \) \] 
 \ee
 and hence has poles at $3,5,7,\dots$. 
 
In order to have agreement with (\ref{tab:UV1}) $I/c_d$ must have a zero for $d=2$ at all values of the parameters. Indeed this is found to be the case. Inside the $\lam=0$ cone this zero is  a consequence of the identity that the sum of angles of a (Euclidean) triangle is $\pi$.  On the cone this will be seen in the next subsection below  (\ref{lam_poles}). 
Finally outside the cone the $d=2$ zero is a result of \be 
I/c_d \propto \log \prod_{i=1}^3 \frac{s^i + \sqrt{\lam}/2}{s^i - \sqrt{\lam}/2} = \log \frac{A(\{s^i\}) + 0 \cdot \sqrt{\lam}}{A(\{s^i\}) - 0 \cdot \sqrt{\lam}} =0
\ee
where $A(\{s^i\})=s^1s^2 s^3 + \lam (s^1+s^2+s^3)/4$ and (\ref{lam_s}) was used to obtain the cancellation of the coefficient of $\sqrt{\lam}$ in both numerator and denominator.

This zero can be represented as a negative multiplicity inside the series of poles, namely $2^{-1},\, 3,5,7,\dots$.

The following table summarizes the UV pole structure as seen in the $s$ plane \be \begin{array}{cc}
\mbox{component } & \mbox{pole series }  \\
 \hline 
 c_d		 	& 2, 4^2, 6^2, 8^2,\dots	 \\
  I/c_d	 	& 2^{-1}, 3,5,7 ,\dots  \\
\\ \hline 
\mbox{Total } & 3,\, 4^2, 5,\, 6^2, \dots  \\
\hline
\end{array}
\label{tab:UV2}
\ee

The total series fully agrees with that obtained from the $l\, /\, \al$ plane (\ref{tab:UV1}). It is notable that the division of the poles between the various sources works rather differently. 

\subsection*{The $\lam=0$ locus}

On the $\lam=0$ cone the expression for $I$ (\ref{result}) simplifies considerably and becomes \be
 I |_{\lam=0} = \frac{c_d}{d-3} \sum_{i=1}^3 (s^i)^{d-3} ~.
\label{alg_soln_revised}
\ee

Comparing with the available expression at $\lam=0$ (\ref{AlgLocSoln}), the current expression is seen to be equivalent and indeed simpler. This simplification relies on the identity \be
\frac{x_1^2 + x_2^2 - x_3(x_1 + x_2)}{x_1 + x_2 + x_3} = s^3 \mod \lam
\label{div}
\ee
and its two permuted versions. To confirm it one may attempt $ x_1^2 + x_2^2 - x_3(x_1 + x_2) = a(x) (x_1 + x_2 + x_3) + b\, \lam$ where $a(x),\, b$ are yet unknown.  Setting $x_3=-x_1-x_2$ allows to determine $b$ to be $b=1/2$. Now $a(x)$ can be determined to be $a(x)=s^3$, thereby proving (\ref{div}).  The simplification was not manifest in (\ref{AlgLocSoln}) because the divisibility occurs only $\mod \lam$ .

The pole structure of (\ref{alg_soln_revised}) relative to the generic one (\ref{tab:UV1},\ref{tab:UV2}) is given by \be
d=5^{-1},7^{-1},9^{-1}, \dots
\label{lam_poles}
\ee
namely, the poles at $5,7,9,\dots$ disappear locally at $\lam=0$. Note that at $d=2$ the pole of $c_d$ is still cancelled against the zero of $I/c_d = \sum_{i=1}^3 (s^i)^{d-3} /(d-3)= -\sum (s^i)^{-1} = \lam/(4\, s^1\, s^2\, s^3)=0$, where the second to last equality uses the expression for $\lam$ in terms of $s^i$ (\ref{lam_s}).

Analyticity. Let us comment on the analyticity of $I(x_1,x_2,x_3)$. The expression (\ref{result}) is analytic for all positive $x_i$. Yet the cone $\lam(x_1,x_2,x_3)=0$ is special as the integral has a pseudo threshold over there, characterized by a solution of the Landau equations such that all momenta are collinear, yet not all of the Schwinger parameters are positive (see e.g. \cite{ItzyksonZuber}). The prototypical example for a pseudo threshold is the bubble diagram (1-loop propagator diagram) with $s=(m_1-m_2)^2$. At a pseudo threshold the diagram is analytic on the physical sheet, yet singularities exist on other sheets, for instance it may be given by a logarithm which produces an analytic function once expanded around its principal value.

\subsection*{IR divergences}

\noindent {\bf A massless propagator}. We have seen how UV divergences result in poles in the $d$-plane which are generic in the parameter space. IR divergences, on the other hand, cause additional $d$-poles to appear when some of the propagators become massless. 

In this part we shall consider the case when a single propagator is massless, for concreteness $x_3=0$ and $x_1 \ge x_2$. In this case $s^2=-s^1=(x_1-x_2)/2,\, \lam=(x_1-x_2)^2$ and the general expression for $I$ (\ref{result}) simplifies to \be
I = -c_d  \int_{s^3}^\infty \(s^2-\lam/4 \)^{d/2-2} ds
\label{I_at_x30}
\ee
where $s^3=(x_1+x_2)/2$.

A parametric evaluation (along the lines of appendix \ref{app:eval_A}) yields another expression in terms of a single integral \be
 I = - 2 \pi^d\, \Gamma(2-d)\, \int_0^1 d\beta \[\beta x_1 + (1-\beta) x_2\]^{d-3} \[\beta (1-\beta)\]^{1-d/2}
\ee 
The two last expressions must be equivalent, though it is not apparent. 

The pole structure of (\ref{I_at_x30}) differs from the generic one (\ref{tab:UV2})  by \be
d=2
\label{1massless_poles}
\ee
namely, the addition of a pole at $d=2$. This pole represents an IR divergence, as one might expect by inspecting the definition of the integral (\ref{def:I}).  The associated residue, that is, the coefficient of $1/(d-2)$  is \be
\mbox{Res}|_{d=2} = 2 \pi^2 \frac{1}{x_1-x_2} \log \frac{x_1}{x_2}
\label{res2}
\ee

We would like to make a few comments. First, note that the general expression (\ref{result}) is a sum of three terms, each of the same form as the one for massless $x_3$, namely (\ref{I_at_x30}). Secondly,  the pole at $d=2$ appears since $I/c_d$ no longer has a zero, which it had for generic values of the masses. Presumably this is a consequence of a change in the order of limits. Thirdly, we note that the residue (\ref{res2}) is positive for all values of $x_1,x_2$. Finally, when considering the limits of integration (\ref{def:contour}) and their influence on the pole structure we can identify $\infty$ as being responsible for UV divergences and hence it can be called a {\it UV limit} while by the same logic $s^i=\sqrt{\lam/4}$ is an {\it IR limit}.

\presub {\bf Two massless propagators}. This case is given by (\ref{I_A}) and derived in appendix \ref{app:eval_A}. The pole structure relative to the generic one (\ref{tab:UV2}) is given by \be
d=\dots, -4,\, -2,\, 0,\, 2^2
\label{poles_A}
\ee
Hence, the additional poles relative to the previously discussed case of a single massless propagator (\ref{1massless_poles})  are at $d=\dots, -4,\, -2,\, 0,\, 2$ and describe IR divergences.



\section{Summary and discussion}
\label{sec:discussion}

This paper solves the SFI equations for the diameter diagram, fig. \ref{fig:diam}.  The resulting expression for the value of the diagram for the most general masses and space-time dimension is given in (\ref{result}-\ref{def:contour}). 

Given that the general value was already determined in 1992 by Ford Jack and Jones  \cite{FJJ1992} and by Davydychev and Tausk  \cite{DavydTausk1992} I would like to explain the value added by the current paper.

First, this paper demonstrates the utility of the SFI equations, whose solution determines the dependence of the integral on all of its parameters (since the co-dimension of the SFI group is 0), up to a determination of a base point in parameter space (two massless propagators) as well as either a boundary conditions at $\lam=0$ or a second base point at equal masses. 

Secondly, we can compare the expression in this paper with the literature. Davydychev and Tausk (DT) used the Mellin-Barnes transform\footnote{
The Mellin-Barnes (MB) transform of massive propagators was introduced in \cite{BoosDavyd1990} -- see also references therein.}
 and obtained an expression in terms of the hypergeometric function -- see eq.s (4.5,4.12) there. Ford Jack and Jones (FJJ)  solved differential equations through the method of characteristics and obtained an integral expression - see eq.s (4.14,4.16) there. This expression is equivalent to that of DT,  yet simpler in the sense that it is clear how to pass from the FJJ expression to the DT expression, but not vice versa. Comparing with the expression in this paper one finds that it has the same integrand as the FJJ expression, yet the integration limits (or integration contour) are different, though equivalent. The current contour has the advantage of manifest continuity at $\lam=0$. In addition, inspired by \cite{Davyd1999,DavydDelbourgo1997} this paper provides a geometrical interpretation of elements of the result in section \ref{sec:triangle}.

Thirdly, we compare methods of derivation.  The SFI method defines a set of differential equations, (\ref{dimeq}-\ref{sl2eq}), and uncovers its underlying geometry in parameter space. The individual differential equations are the same as those obtained in the DE method, and employed by FJJ, yet SFI stresses the equation set as a whole. In fact, FJJ do not state the equation set, even though several of the equations are mentioned and solved. 

This paper supplies three derivations with the hope that different methods may be useful in future analyses of other diagrams. The first derivation shares the same method as FJJ while the other are novel and employ the $S_3$ permutation symmetry or partial invariants, which are especially inspired by the SFI perspective.
 
Fourth, we analyze the UV and IR divergences for this diagram in general dimension in section \ref{sec:diverg}. 

In summary,  we demonstrate the utility of the SFI equations for evaluating the diameter diagram. Our treatment is close to that of FJJ, yet the current expression for the result differs a bit as discussed above, it is interpreted geometrically, novel derivations are provided and an analysis of divergences is given.

\subsection*{Acknowledgments}

This research started in summer 2015 as an ``Amirim'' undergraduate project of Erez Urbach (supervised by the author), and was developed significantly during a stay in the Italian Alps in August 2017.

It is a pleasure to thank Erez Urbach for collaboration during early stages of this work, as well as Philipp Burda and Ruth Shir for collaboration on related projects and for comments on a preliminary presentation. 

This research was partly supported by the ``Quantum Universe'' I-CORE program of the Israeli Planning and Budgeting Committee (since November 2015), by the Einstein Research Project ``Gravitation and High Energy Physics" funded by the Einstein Foundation Berlin (till August 2016) and by the Israel Science Foundation grant no. 812/11 (till September 2016).

\appendix

\section{Evaluation at $m_2=m_3=0$}
\label{app:eval_A}

\begin{figure}
\centering \noindent
\includegraphics[width=4cm]{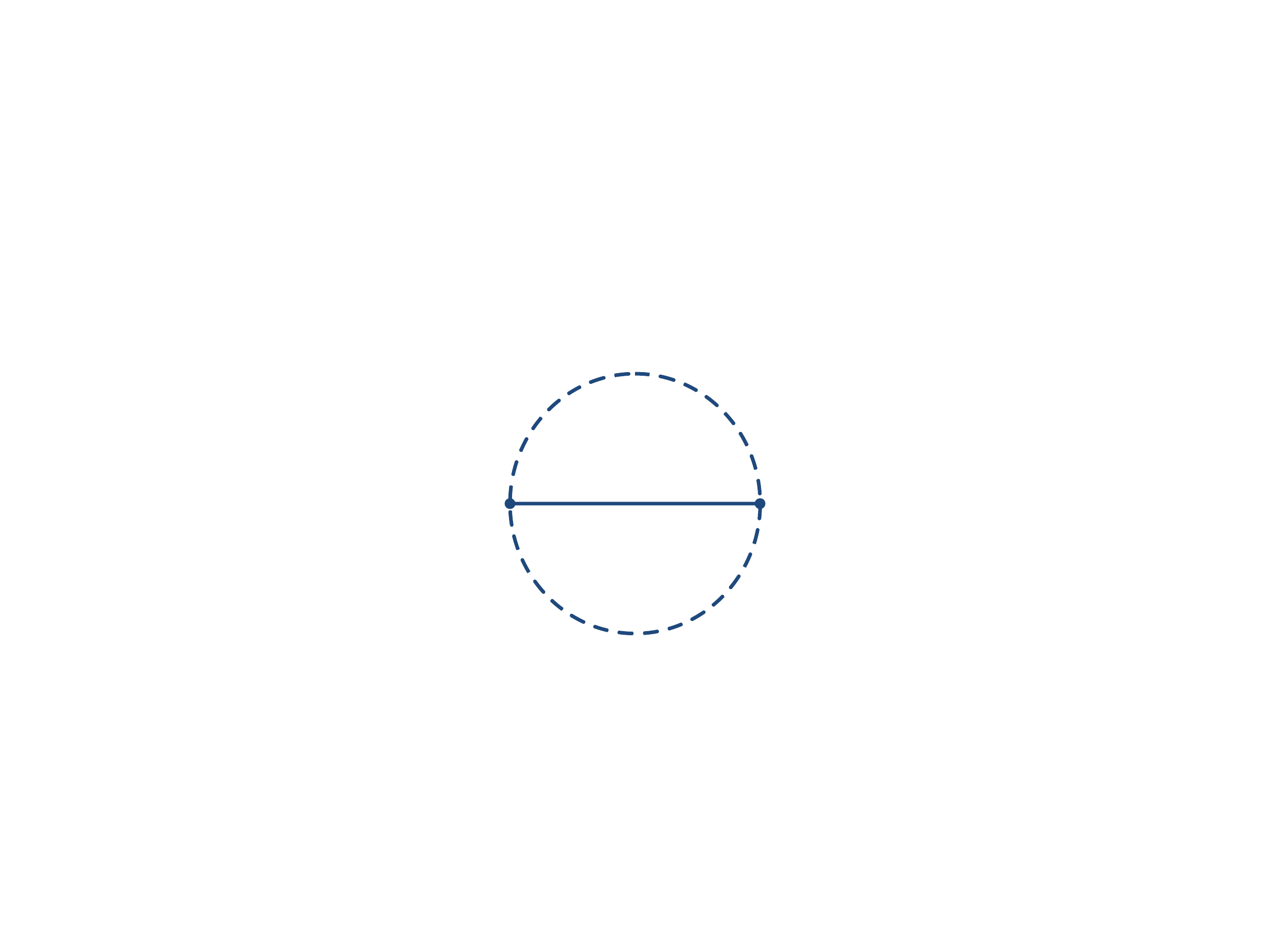} 
\caption[]{The diameter diagram with two massless propagators denoted by the dashed lines.}
\label{fig:diamA}
\end{figure}

In this appendix we detail the evaluation of the diameter diagram when two of the propagators are massless, namely $x_1=\mu,\, x_2=x_3=0$, see also figure \ref{fig:diamA}.

The computation is performed in a rather standard way. First we analytically continue to the Euclidean integral \be
I_A = (+i)^L\, (-)^P I_A^E = +I_A^E
\ee
where $I_A$ is the Lorentzian integral under study, $L=2$ is the number of loops, $P=3$ is the number of propagators and $I_A^E$ is the Euclidean integral which we proceed to define and compute \bea
 I_A^E  &:=& \int \frac{d^dl_1\, d^dl_2}{ \((l_1)^2 + \mu\)\, (l_2)^2\, (l_1+l_2)^2} = \non
	&=& \int_0^\infty d\al^1 d\al^2 d\al^3\, \exp(-\al^1 \mu) \times \non
	&& \times \int d^dl_1\, d^dl_2\, \exp \( - \[ (\al^1+\al^3) (l_1)^2 + 2 \al^3\, l_1 \cdot l_2 + (\al^2+\al^3) (l_2)^2\] \)= \non
	&=& \pi^d\, \int_0^\infty d\al^1 d\al^2 d\al^3\, \exp(-\al^1 \mu) \[\al^1(\al^2+\al^3) + \al^2 \al^3 \]^{-d/2} = \non
	&=& \pi^d\, \mu^{d-3}\, \Gamma(3-d)\,  \int_0^\infty d\tal^2 d\tal^3\, \[ \tal^2 + \tal^3 + \tal^2 \tal^3 \]^{-d/2} = \non
	&=& \pi^d\, \mu^{d-3}\, \Gamma(3-d)\, \frac{1}{d/2-1} \int_0^\infty d\tal^2 \frac{(\tal^2)^{1-d/2}}{1+\tal^2} = \non
	&=&  \pi^d\, \mu^{d-3}\, \Gamma(2-d)\, \frac{2 \pi}{\sin (\pi d/2)} ~~. 
	\eea
The first line is the definition of the Euclidean integral, in the 2nd the Schwinger parameters are introduced, next we integrate over the loop momenta and the Kirchhoff-Symanzik polynomial appears.  In passing to the 4th equality we make use of $x_2=x_3=0$ to define $\tal^2:=\al^2/\al^1,\, \tal^3:=\al^3/\al^1$ and integrate over $\tal^1$. Next we integrate over $\tal^3$. In passing to the last line we use  $\Gamma(3-d)/(2-d) = \Gamma(2-d)$ as well as \bea
 && \int_0^\infty d\tal^2 \frac{(\tal^2)^{1-d/2}}{1+\tal^2} = \int_0^1dy\, y^{d/2-2} \(1-y \)^{1-d/2} = \non
 &=& B\(d/2-1,2-d/2\) = \frac{\pi}{\sin \pi d/2} 
 \eea
 where we changed variables to $y:=1/(1+ \tal^2)$, and used the gamma function identity (\ref{Gamma_reflection}) to simplify the beta function.
 
Summarizing \be
  I_A = \pi^d\, \mu^{d-3}\, \Gamma(2-d)\, \frac{2 \pi}{\sin (\pi d/2)} ~~. 
\label{I_A2}
\ee
  
\presub {\bf Test}.  We test the overall sign by noting that the integrand in the definition of $I_E$ is positive, and in the range of dimensions $2<d<3$ there are neither UV nor IR divergences, and hence $I_E$ should be positive, as indeed (\ref{I_A2}) confirms.
 
\section{Comparison inside the cone}
\label{app:FJJ_inside}

Here we detail how comparison with \cite{FJJ1992} allows to determine the mass-universal constant inside the cone given by the contour (\ref{def:cont2}). This reference uses the method of characteristics to solve certain differential equations (which are included in the SFI set), at least outside the cone. Regarding the region inside the paper states that ``it is possible to derive'' eq. (4.16) which in the current notation reads \be
I = -I(\sqrt{-\lam},0,0)\, \sin \frac{\pi d}{2} + c_d \[ \sum_{i=1}^3 \int_0^{s^i} \] \(s'^2 -\lam/4\)^{d/2-2}\, ds' ~.
\label{FJJ_inside}
 \ee
 
The first term can be evaluated through (\ref{I_A})  to \be 
 I(\sqrt{-\lam},0,0)\, \sin \frac{\pi d}{2} = \pi^d\, \sqrt{-\lam}^{d-3}\, 2 \pi\, \Gamma(2-d) 
 \ee
On the other hand the same expression equals also \be
 c_d \int_0^\infty  \(s'^2 -\lam/4\)^{d/2-2}\, ds'
\ee
where the computation uses the integral $\int_0^\infty (s^2+1)^{d/2-2} ds=B\(\half,\frac{3-d}{2} \) /2$.

Together this implies that (\ref{FJJ_inside}) can be represented by the contour \be
\int_{C_{FJJ}} :=  \sum_{i=1}^3 \int_0^{s^i} - \int_0^\infty 
\ee
which equals \be
   \sum_{i=1}^3\( \int_0^\infty - \int_{s^i}^\infty\) - \int_0^\infty = \int_{-\infty}^{+\infty} -  \sum_{i=1}^3 \int_{s^2}^\infty \equiv \int_C ~,
 \ee
namely the contour (\ref{def:cont2}), thereby completing the computation.

\section{Gamma function identities}
\label{app:Gamma}

In this short appendix we collect identities of Gamma functions \bea
x \Gamma (x) = \Gamma (x+1) \label{Gamma_defining} \\
 \Gamma(x)\, \Gamma(1-x) &=& \frac{\pi}{\sin \pi x} \label{Gamma_reflection} \\
\Gamma(x)\, \Gamma(x+\half) &=& \frac{\sqrt{\pi}}{2^{2x-1}}\, \Gamma(2x) \label{Gamma_doubling} 
\eea

\bibliographystyle{unsrt}

\begin{thebibliography}{99}

 \bibitem{SFI}  
  B.~Kol,
  ``Symmetries of Feynman integrals and the Integration By Parts method,''
  arXiv:1507.01359 [hep-th].

\bibitem{ChetyrkinTkachov1981} 
  K.~G.~Chetyrkin and F.~V.~Tkachov,
  ``Integration by Parts: The Algorithm to Calculate beta Functions in 4 Loops,''
  Nucl.\ Phys.\ B {\bf 192}, 159 (1981).
%

\bibitem{DE}
  A.~V.~Kotikov,
  ``Differential equations method: New technique for massive Feynman diagrams calculation,''
  Phys.\ Lett.\ B {\bf 254}, 158 (1991). \\
  E.~Remiddi,
  ``Differential equations for Feynman graph amplitudes,''
  Nuovo Cim.\ A {\bf 110}, 1435 (1997)
  [hep-th/9711188]. \\
  T.~Gehrmann and E.~Remiddi,
  ``Differential equations for two loop four point functions,''
  Nucl.\ Phys.\ B {\bf 580}, 485 (2000)
  [hep-ph/9912329].

\bibitem{SmirnovBooks} 
  V.~A.~Smirnov,
  ``Feynman integral calculus,''
  Berlin, Germany: Springer (2006). 
  V.~A.~Smirnov,
  ``Analytic tools for Feynman integrals,''
  Springer Tracts Mod.\ Phys.\  {\bf 250}, 1 (2012).

\bibitem{VacSeagull} 
  P.~Burda, B.~Kol and R.~Shir,
  ``Vacuum seagull: Evaluating a three-loop Feynman diagram with three mass scales,''
  Phys.\ Rev.\ D {\bf 96}, no. 12, 125013 (2017)
  doi:10.1103/PhysRevD.96.125013
  [arXiv:1704.02187 [hep-th]].
   
\bibitem{bubble} 
  B.~Kol,
  ``Bubble diagram through the Symmetries of Feynman Integrals method,''
  arXiv:1606.09257 [hep-th].

\bibitem{FJJ1992} 
  C.~Ford, I.~Jack and D.~R.~T.~Jones,
  ``The Standard model effective potential at two loops,''
  Nucl.\ Phys.\ B {\bf 387}, 373 (1992)
  Erratum: [Nucl.\ Phys.\ B {\bf 504}, 551 (1997)]
  doi:10.1016/0550-3213(92)90165-8, 10.1016/S0550-3213(97)00532-4
  [hep-ph/0111190].

\bibitem{DavydTausk1992} 
  A.~I.~Davydychev and J.~B.~Tausk,
  ``Two loop selfenergy diagrams with different masses and the momentum expansion,''
  Nucl.\ Phys.\ B {\bf 397}, 123 (1993).

\bibitem{Davyd1999} 
  A.~I.~Davydychev,
  ``Explicit results for all orders of the epsilon expansion of certain massive and massless diagrams,''
  Phys.\ Rev.\ D {\bf 61}, 087701 (2000)
  doi:10.1103/PhysRevD.61.087701
  [hep-ph/9910224].

\bibitem{DavydDelbourgo1997} 
  A.~I.~Davydychev and R.~Delbourgo,
  ``A Geometrical angle on Feynman integrals,''
  J.\ Math.\ Phys.\  {\bf 39}, 4299 (1998)
  doi:10.1063/1.532513
  [hep-th/9709216].

\bibitem{DavydychevTausk95magic} 
  A.~I.~Davydychev and J.~B.~Tausk,
  ``A Magic connection between massive and massless diagrams,''
  Phys.\ Rev.\ D {\bf 53}, 7381 (1996)
  doi:10.1103/PhysRevD.53.7381
  [hep-ph/9504431].
  
 \bibitem{locus} 
  B.~Kol,
  ``The algebraic locus of Feynman integrals,''
  arXiv:1604.07827 [hep-th].

\bibitem{Davyd2016} 
  A.~I.~Davydychev,
  ``Geometrical splitting and reduction of Feynman diagrams,''
  J.\ Phys.\ Conf.\ Ser.\  {\bf 762}, no. 1, 012068 (2016)
  doi:10.1088/1742-6596/762/1/012068
  [arXiv:1605.04828 [hep-th]].

\bibitem{ItzyksonZuber}
C.~Itzykson adn J.-B.~Zuber,
``Quantum Field Theory,''
McGraw-Hill (1980), section 6-3.

\bibitem{BoosDavyd1990} 
  E.~E.~Boos and A.~I.~Davydychev,
  ``A Method of evaluating massive Feynman integrals,''
  Theor.\ Math.\ Phys.\  {\bf 89}, 1052 (1991)
  [Teor.\ Mat.\ Fiz.\  {\bf 89}, 56 (1991)].
  doi:10.1007/BF01016805

\end{thebibliography}

\end{document}